\documentclass[reprint,amsmath,amssymb,aps]{revtex4-2}

\usepackage{graphicx}  
\usepackage{color}
\usepackage{dcolumn}% Align table columns on decimal point
\usepackage{bm}% bold math
\begin{document}

\preprint{AIP/123-QED}

%\documentclass[a4paper, amsfonts, amssymb, amsmath, reprint, showkeys, nofootinbib, https://www.overleaf.com/project/653b8bcef7c27317048f2ac7twoside]{revtex4-1}
%\usepackage[english]{babel}
%\usepackage[utf8]{inputenc}
%\usepackage[colorinlistoftodos, color=green!40, prependcaption]{todonotes}
%\input{preamble}
%\usepackage[pdftex, pdftitle={Article}, pdfauthor={Author}]{hyperref} % For hyperlinks in the PDF
%\setlength{\marginparwidth}{2.5cm}
%\bibliographystyle{apsrev4-1}
%\begin{document}
\title{Strongly interacting bosons in 1D disordered lattice: \\ phase coherence of distorted Mott phases}
\author{Barnali Chakrabarti$^{1,2}$, Arnaldo Gammal$^{2}$, Luca Salasnich$^{3}$}
%\email{bchakrab@ictp.it}
\affiliation{$^1$Department of Physics, Presidency University, 86/1   College Street, Kolkata 700073, India.}
\affiliation{$^2$Instituto de Física, Universidade de São Paulo, CEP 05508-090, SP, Brazil.}
\affiliation{$^3$Dipartimento di Fisica ``Galileo Galilei", Universita di Padova, INFN Sezione di Padova, and CNR-INO Unita di Sesto Fiorentino, Italy}
\date{\today} % Leave empty to omit a date

\begin{abstract}
We explore the consequences of disorder on phase coherence in the Mott insulator phases in an optical lattice. Few bosons with contact interaction in small optical lattice can feature varieties of insulating phases - weakly interacting Mott in deep lattice, maximally fragmented and strongly interacting Mott in intermediate lattice, weak Mott with double filling and intra-well coherence, fermionized Mott with strong intra-well coherence. Utilization of the multiconfigurational time dependent Hartree method for bosons (MCTDHB) to solve the many-boson Schr\"odinger equation, facilitates to understand the microscopic effect of disorder on the different kinds of Mott phases in the primary lattice. The many-body properties are analyzed by distinct measures of the reduced one-body density in real and momentum space, fragmentation, order parameter, variance of spatial single-shot measurements, compressibility and the Glauber normalized correlation functions. We find very complex competition of localization due to disorder and Mott correlation. We observe distinct response of four different Mott phases in the disordered lattice. When the weakly interacting Mott exhibits the Bose Glass phase, the strongly correlated and fully fragmented Mott exhibits leakage, melting and central localization. In contrast, weak Mott with double filling which is a spatially separated dimer in each site, exhibits only some dissociation of intra-well coherence and assist in the development of inter-well coherence in the presence of strong disorder. For the fermionized Mott, when the density in each well is fragmented, strong disorder interfere with the intra-well correlation and the characteristic dip in each site starts to disappear leading to simple Mott localization with a pair of bosons.  
  
\end{abstract}


\maketitle

\section{Introduction} \label{sec:intro}
The localization of non-interacting systems in a disordered medium was predicted by Anderson in his seminal work~\cite{Anderson}. Since then the study of interplay between disorder and interaction becomes the most challenging research in strongly correlated systems and is more intriguing in reduced dimension. It exhibits fascinating phenomena: many-body localization~\cite{Bloch:2019}, quantum phases~\cite{Giamarchi:1988,Fisher:1989} and collective Anderson localization~\cite{Lugan:2007,Lugan:2011,Samuel:2015}. Among the different setups to realize the localization effect, one-dimensional (1D) bichromatic quasi-periodic optical lattice has gained particular interest and was used in the experiment of Roati {\it et al.}~\cite{Roati:2008}. The spatial ordering in 1D quasi-periodic potential is intermediate between periodic and random potential~\cite{Sokoloff:1985,Thouless:1983}. The 1D system subjected to the quasi-periodic potential is modeled by Aubry-Andr\`e (AA) potential~\cite{Aubry:1980} which exhibits interesting property--a transition from the extended to the localized states~\cite{Paliencia:2005,Paliencia:2007,Fallani:2005}. However beyond the realization of AA model, quasi-periodic disorder exhibits more intriguing properties~\cite{Palencia:2019,Fangzhao:2018,Saurabh:2021,Sarma:2011,Sarma:2015,Masudul:2013, Wang:2020,Fangzhao:2021,Roy:2021,Anna:2021}. The intermediate consequence of disorder on interacting bosons in 1D optical lattice is the emergence of new quantum phases. The localized Mott-insulator (MI) phase and the delocalized superfluid (SF) phase are intercepted by the presence of a third phase--the Bose glass (BG) phase~\cite{Giamarchi:1988,Fisher:1989,Fallani:2005,Palencia:2019,Pasienski:2010,Yu:2012,Ronen:2012}. The BG phase is characterized as an insulating phase having finite compressibility. The phase diagram of 1D disordered bosons has also been theoretically studied ~\cite{Lugan1:2007,Luca:2009}. Controlled quasi-periodic potential with ultracold atoms can be realized experimentally~\cite{Modugno:2010,Palencia:2010}. It is interesting to characterize how the tailored weak disorder can be engeneered to induce the BG transition or to distabilize the SF phase~\cite{Cazalilla:2011}.    

Theoretically the phases of interacting bosons in the disordered lattice have been investigated in the frameworks of quantum Monte Carlo simulation~\cite{Palencia:2020,Pinaki:NJP, Richard:1991,Moore:2018,yao2024motttransitionlieblinigergas}, mean-field theories~\cite{Adhikari:2009,Damski:2003,Lewenstein:2005,Kuhn:2005} and DMRG~\cite{Zwerger:1999}. However the Bose-Hubbard model (BHM) is considered as the paradigmatic model to describe the phases  in lattice~\cite{Jaksch:1998,RevModPhys.80.885,Fisher:1989,Pinaki:2007,Esslinger:2004,Noor:2023}.  In the BHM model, a weakly interacting Bose gas in an optical lattice of moderate depth corresponds to the superfluid phase. For the unit filling case, the SF phase persists as long as the tunneling coupling $(J)$ of the Bose-Hubbard Hamiltonian is more than the interparticle interaction $(U)$ which corresponds to long-range coherence. Otherwise the superfluid phase persists for arbitrary 
repulsion U. When the interaction strength is large compared to the tunneling coupling--long-range coherence is lost and localization happens in each site, the many-body state makes a transition to the Mott insulating (MI) phase. 
Even though, the Bose-Hubbard model can aptly describe
the above SF to MI transition in lattices, its regime of validity
is highly restricted for site localized Wannier states~\cite{Astrakharchik:2016}. The SF to MI transition is determined by the relative strength of $U$ and $J$, not by their independent tuning. The single parameter $\frac{U}{J}$ determines the phase diagram irrespective of how the ratio is achieved, i.e, the change in interaction is equivalent to the change in lattice depth. 

In this article we go beyond the BH model. It has been demonstrated
that a general quantum many-body description, valid at all
interaction strengths, is necessary for the regime beyond the
Bose-Hubbard model. One such theory is the multiconfigurational time-dependent Hartree (MCTDH) method~\cite{Meyer:2000,Sascha:2010}. In the present
paper, we solve the full many-body Schr\"odinger equation at a
high level of accuracy by using the multiconfigurational time-
dependent Hartree method for bosons (MCTDHB)~\cite{Alon:2007,Alon:2008,Streltsov:2006,Streltsov:2007,Lode:2016,Fasshauer:2016,Lode:2020}. We cover the different interaction regime starting from weak interaction to the fermionized limit for few boson ensemble with integer filling in a finite 1D disordered lattice. Independent control of interaction parameter, lattice depth and filling factor facilitate to configure {\it four} different kinds of clean Mott phases (without disorder) exhibiting entire range of weak to strong intra-well correlations. These are :
i) weakly interacting Mott in deep lattice with unit filling, ii) strongly interacting Mott in lattice of moderate depth with unit filling, iii) doubly filling and weakly interacting Mott in lattice of moderate depth, iv) fermionized Mott--strongly interacting Mott with double filling which exhibits strong fragmentation in each lattice.

All the four initial phases are scrutinized under the presence of incommensurate second lattice which introduces disorder. Instead of studying the phase diagram which is mostly studied in earlier works, we will rather focus to understand the interplay between Mott localization and localization due to disorder. The investigation of phase coherence for strongly interacting Mott phase with unit filling and fermionized Mott phase with double filling in disordered lattice, which is the beyond Bose-Hubbard physics; are the main motivations of the present work.

We present numerically exact results by employing MCTDHB. We will use the open source implementation of
MCTDHB in the MCTDH-X software~\cite{Lin:2020,MCTDHX}.
The basic key quantities addressed are : a) reduced one-body density in real space, which exhibits the effect of localization introduced by the disorder, b) zero-momentum peak in the one-body momentum distribution to exhibit the competition between Mott localization and localization due to disorder, c) fragmentation, order parameter, compressibility and the variance of spatial single-shot measurements to distinguish the BG phase from MI phase, d) one-body and two-body correlation functions--the distinct measures to understand the response of strongly interacting Mott phase in the disordered lattice. 

Our general observations are as follows: 1) The weakly correlated Mott with unit filling enters to the BG phase on switching on the disorder and remains as the BG phase throughout. 2) The strongly interacting fully fragmented Mott with unit filling does not exhibit any BG phase, rather the strongly correlated Mott lobes exhibit a complex pathway of leakage-melting-central localization.  
3) Weakly interacting Mott with double filling exhibits intra-well two-body coherence. In the presence of strong disorder, enhancement of central localization happens. With stronger disorder, depletion in the diagonal coherence and enhancement of off-diagonal correlation are observed. 4) Fermionized Mott exhibits strong on-site repulsion between the pair of bosons which leads to fragmentation of each pair into two orbitals in the same well. In the presence of second lattice, the disorder competes with the on-site repulsion between the pair of bosons and the characteristic dip in each well becomes gradually disappeared. Much stronger disorder leads to simple Mott localization with a pair of bosons in each well. 

The paper is organized as follows. The model and theoretical approach is presented in Section II, Section III presents the basic measures. Section IV describes the initial states. The numerical results are presented in Sec V distributed over three subsections.
Section VI concludes the summary.

\section{Model and Theoretical approach}
We consider $N$-bosons with repulsive contact interaction subjected to a 1D quasi-periodic potential $V(x)$. The Hamiltonain is
\begin{equation}
    H=\sum^N_{i=1}\left[ -\frac{\hbar^2}{2m}\frac{\partial^2}{{\partial x_i}^2}+V(x_i)\right]+\lambda \sum_{i<j}
\delta(x_i-x_j)  
\end{equation}
where $m$ is the particle mass and $\lambda$ is the two-body interaction strength. The quasi-periodic potential is 
\begin{equation}
    V(x) = V_1 \sin^2(k_1 x) + V_2 \sin^2 (k_2 x)
\end{equation}
where $k={\pi}/{d}$ is the wave vector and $d$ is the periodicity of lattice. $V_1$ and $V_2$ measure the height of the lattices in units of recoil energies $E_{r_1}= \frac{\hbar^2 k_1^2}{2m}$ and $E_{r_2}= \frac{\hbar^2 k_2^2}{2m}$ respectively. $k_1$ and $k_2$ are chosen such that the experimental wavelengths $\lambda_1$=1032 nm and $\lambda_2$ = 862 nm become $\lambda_1$ $\simeq$ 1 and $\lambda_2$ $\simeq$ 0.86 in dimensionless units. For quasi-periodic lattice $\frac{k_2}{k_1} =r$, $r$ is irrational. Here we use $r \simeq 1.1972....$ close to the experiments of the Refs.~\cite{Roati:2008,Billy:2008}.
$V_1$ is the primary lattice and $V_2$ introduces the disorder.
For a finite optical lattice we impose hard wall boundaries at the appropriate positions.
For unit filling, we investigate maximum nine-well set up $(S=9)$ with $N=9$ bosons and computation is done with $M=9$ orbitals. For the double filling case we use three-well set up with six bosons and computation is done with $M=12$ orbitals. These few-atom systems are handled numerically exactly with very high precision even for very strong interaction. This finite ensemble can be considered as a fundamental building blocks of many-body systems from a bottom-up research. All the parameters are tunable in the present day experiments. The interaction strength $\lambda$ is chosen to cover the entire region from weakly interacting to fermionized limit. As previously discussed, for the reasons of universality and computational convenience we rescale the Hamiltonian in units of the recoil energy $E_{r}= \frac{\hbar^2 k^2}{2m}$, setting $\hbar$= $m$ = $k$ = $1$. The length and time units are given in terms of $k^{-1}$ and $\frac{\hbar}{E_r}$ respectively. All quantities are in dimensionless units throughout. 

%\section{Many-body method}
The MCTDHB method uses a multiconfiguration time adaptive ansatz to solve the time dependent Schr\"odinger equation
\begin{equation}
    \hat{H}\vert {\Psi} \rangle = i \partial_t \vert {\Psi} \rangle
\end{equation}
where $|\Psi>$ represents the many-body state of interacting indistinguishable bosons. The ansatz for MCTDHB wave function is 
$\vert {\Psi(t)} \rangle=\sum_{\bar{n}}C_{\bar{n}}(t) \vert{\bar{n},t} \rangle$, $\vert{\bar{n},t} \rangle$ are the time dependent permanents and $\bar{n}$ is the occupation number, $\bar{n}$ =$ \vert{n_1,n_2,\hdots,n_M}\rangle$. $M$ is the number of time dependent single particle states or the number of orbitals. The time dependent orbitals $\phi_j \left( \vec{r},t \right)$ represent the field operator as
\begin{equation}
    \hat{\Psi} (\vec{r},t) = \sum^{M}_{j=1} \hat{b}_j \phi_j (\vec{r}, t)
\end{equation}
Thus $N_{conf}$=  $ \left(\begin{array}{c} N+M-1 \\ N \end{array}\right)$ determines the number of basis states. In the limit of
infinitely many orbitals, $M  \rightarrow \infty$, the set of permanents
$| n_1 ,n_2 , . . . n_M \rangle$  in Eq. (4) spans the complete $N$-particle Fock
space and MCTDHB becomes exact. The use of optimized time-dependent orbitals leads to very fast numerical convergence. Compared to a time-independent basis, as the permanents are time-dependent, a given degree of accuracy is reached with much shorter expansion~\cite{mctdhb_exp1,mctdhb_exp2}. We also emphasize that MCTDHB is more accurate than exact diagonalization which uses the finite basis and is not optimized. Whereas in MCTDHB, as we use a time adaptive many-body basis set, it can dynamically follow the building correlation due to inter-particle interaction~\cite{Alon:2008,Alon:2007, mctdhb_exact3,barnali_axel} and it has been widely used in different theoretical calculations~\cite{rhombik_jpb,rhombik_pra,rhombik_quantumreports, sangita_sci.rep,rhombik_epjd,rhombik_aipconference,rhombik_pre,rhombik_scipost,PhysRevA.109.063308}. In MCTDHB, the convergence is achieved when the measures become independent of increasing in the number of orbitals and the population in the last orbital is insignificant. The MCTDHB equations are derived from the time dependent variational principle~\cite{variational1,variational2,variational3,variational4} - which results into two sets of coupled equations of motion. A set of linear equations for the time dependent coefficients $\{C_{\bar{n}}(t)\}$ and another set of nonlinear equation for $\{ \phi_j(\vec{r},t), j=1,....M \}$ are derived and solved. In the present work, the self consistent ground states are calculated by imaginary time propagation using the MCTDH-X package~\cite{MCTDHX} and the system is relaxed to the ground state. Thus we omit the time dependence in the evaluation of following observable. 

\section{Quantities of interest}
{\it {Fragmentation and order parameter}}\\

%\bigskip
We quantify the fragmentation using the fraction $F$ of atoms which do
not occupy the lowest eigenstate of the reduced one-body
density matrix. Here, 
%The one-body reduced density matrix is a Hermitian matrix and defined as
%$\rho^{(1)}(x|x')$=$\expval{\hat{\psi}^\dag(x')\hat{\psi}(x)}{\Psi(t)}$= $\sum_{kq} %%%%\rho_{kq} \phi_{k}^{\star}(x^{\prime}) \phi_q(x)$, where the matrix element $\rho_{kq} 
%\expval{\hat{b}_k^{\dagger}\hat{b}_q}{\psi}$ represents the one-body reduced density %matrix. It's diagonal gives the probability density 
%$\rho(x)=\rho^{(1)}(x|x^{\prime}=x)$. The diagonalization of Eq(1) is done by an unitary transformation of the orbitals $\phi_n(x)$ to the natural orbitals $\phi_^{(NO)}(x)$ as 
\begin{equation}
\rho^{(1)}(x,x^{\prime}) = \sum_k \rho_k^{(NO)}\phi_{k}^{*(NO)}(x) \phi_{k}^{(NO)}(x^{\prime}).
\end{equation}
$\phi_{k}^{(NO)}(x)$ are the eigenfunctions of $\rho^{(1)}$ and known as natural orbitals. $\rho_k^{(NO)}$ are the eigenvalues of $\rho^{(1)}$ and known as natural occupations. In the mean-field perspective, when one single natural orbital is occupied, $\rho^{(1)}$ has only a single macroscopic eigenvalue, and the many-body state is condensate and nonfragmented. When $\rho^{(1)}$ has $k$ macroscopically occupied eigenvalues - the system is referred to as $k$-fold fragmented~\cite{onsager}.

We define a  mesoscopic ``order parameter" using the
eigenvalues of the reduced one-body density matrix as
\begin{equation}
\bigtriangleup = \sum_{i} \left(  \frac{n_i}{N} \right) ^2,
\end{equation}
where $n_i$ is the natural occupation in $i^{\rm th}$ orbital. 
For the SF phase, $\bigtriangleup=1$ and only one eigenvalue is nonzero. For the Mott phase, the number of significantly contributing orbitals becomes equal to the number of sites $(S)$ and $\bigtriangleup$ = $\frac{1}{S}$. Thus $\bigtriangleup=1$ and $\bigtriangleup$ = $\frac{1}{S}$ are the two extreme values for the two known phases. We find $\bigtriangleup$ is a good marker to identify the Bose glass phase from the superfluid phase. 
We calculate the compressibility from fluctuation in the momentum distribution $K$= $\beta( \langle n^2 \rangle - \langle n \rangle^2 )$, $\beta $ is chosen as significantly high $=1$.\\
\\
{\it{ Higher order densities and correlation functions}}\\
%\bigskip

The $p$-body density is defined as 
\begin{equation}
  \begin{split}
\rho^{(p)}(x_{1}^{\prime}, \dots, x_{p}^{\prime} \vert x_{1}, \dots, x_{p}) = \langle \psi \vert \hat{\psi}^{\dagger}(x_1^{\prime})\dots \hat{\psi}^{\dagger}(x_p^{\prime}) \\ \hat{\psi}(x_p)\dots \hat{\psi}(x_1) \vert \psi \rangle
\label{pbodydensity2}
\end{split}  
\end{equation}
The diagonal of $p$-body density can be represented as 
\begin{equation}
  \begin{split}
\rho^{(p)}(x_{1}, \dots, x_{p})  = \langle \psi \vert \hat{\psi}^{\dagger}(x_1)\dots \hat{\psi}^{\dagger}(x_p) \hat{\psi}(x_p) \\ \dots \hat{\psi}(x_1) \vert \psi \rangle
\label{pbodydensity3}
\end{split}  
\end{equation}
Where $p=2$ gives the two-body density. 
The coherence of the many-body state is defined through Glauber correlation functions. 
 First-order correlation function is defined as  $g^{(1)}(x,x')=\rho^{(1)}(x|x^{\prime})/ \sqrt{\rho(x)\rho(x')}$ ~\cite{Muller:2006,Glauber:1999}. 
 It is the key measure of coherence and experimentally detectable. If the system is in condensed state, $|g^{(1)}(x,x')|$ is constant for all $(x, x^{\prime})$. Fragmented state is diagnosed by loss of coherence $g^{(1)}(x,x') <1$.
The two-body correlation function is defined as $g^{(2)}(x_1,x_2,x_1^{\prime}, x_2^{\prime})$ = $\rho^{(2)} (x_1,x_2,x_1^{\prime},x_2^{\prime} )/ \sqrt{ \rho(x_1)\rho(x_1^{\prime}) \rho(x_2) \rho(x_2^{\prime})}$. 
The diagonal part is 
$g^{(2)} = |g^{(2)}(x_1,x_2, x_1^{\prime}=x_1, x_2^{\prime}=x_2 )|$. $g^{(2)}$ $\neq 1$, indicates partial coherence, the two-body density is not a product of one-body densities. Thus higher order Glauber correlation function provides a spatially resolved measure of coherence. 
\section{Initial Mott configurations}
We solve the many boson Schr\"odinger equation using MCTDHX by propagating in the imaginary time. We explore the many-body phases in the quasi-periodic lattice. In the present computation, we consider a finite-size ensemble which are the best possible choice available in simulation as we need to consider huge number of orbitals to attain the convergence in the correlation measures. However, for such small set up, the ground state possesses correlation properties that are analogous to the macroscopic phases. 
\begin{figure}
    \centering
    \includegraphics[width= 0.4\textwidth, angle = 270]{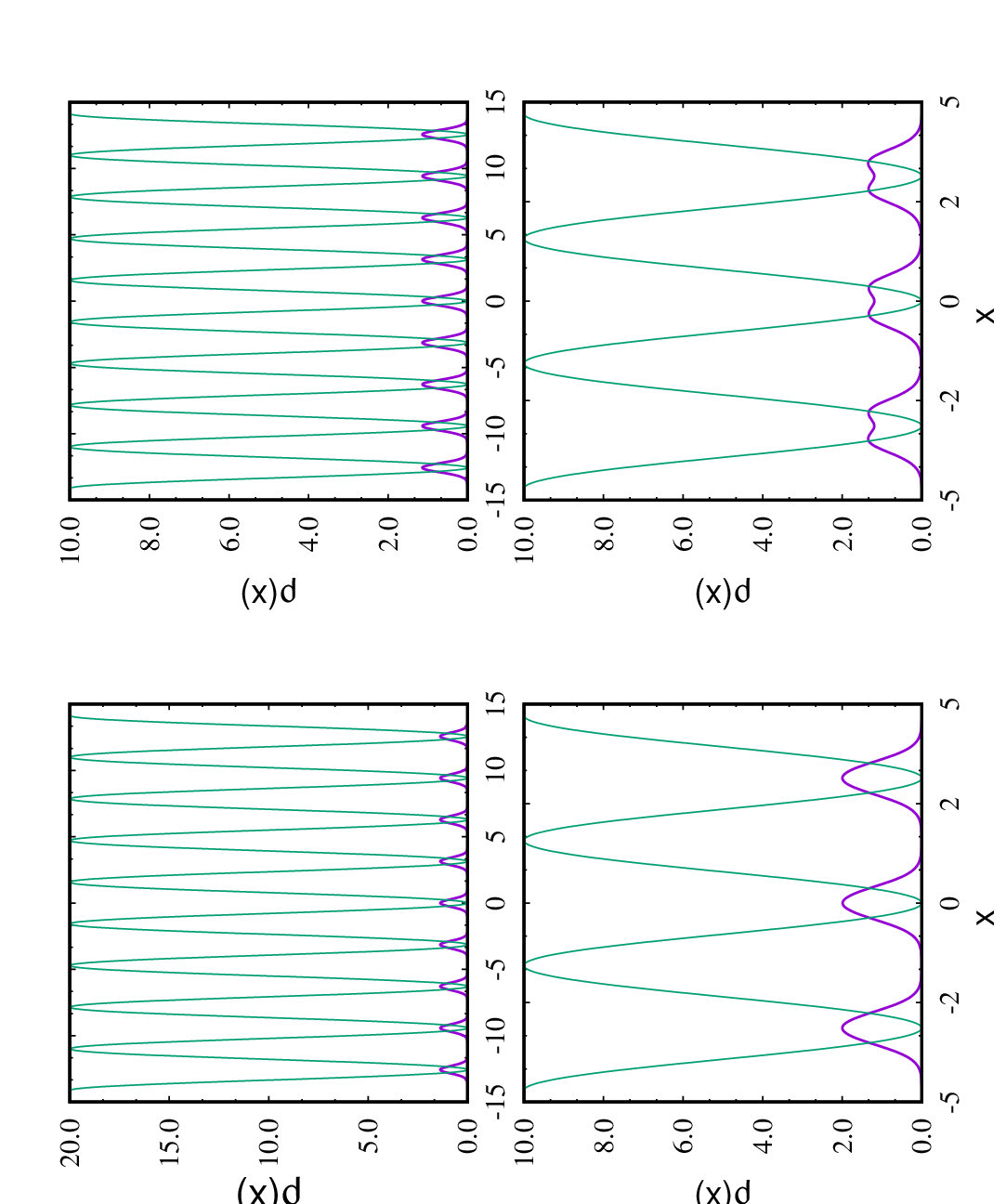}
    \caption{The one-body density $\rho(x)$ (violet color) for four different Mott phases in the primary lattice. The top panel is for unit filling with $N=9$ bosons in $S=9$ lattice sites. Left panel corresponds to weak Mott phase in deep lattice with $\lambda=0.5$, $V_1=20.0$, the right panel is for strong Mott with $\lambda=5.0$ and intermediate lattice of $V_1=10.0$. Computation is done with $M=9$ orbitals. The corresponding potential is drawn in green color. The bottom panels are for double filling with $N=6$ bosons in $S=3$ lattice sites and lattice depth $V_1=10.0$. The left panel corresponds to weak interaction $\lambda = 2.0$, six bosons are equally distributed in three wells. The right panel corresponds to fermionized limit when $\lambda$ is set to $10.0$. The onset of fermionization is mimicked by the dip in each well as each lattice is occupied by two strongly interacting bosons. The lattice potential is drawn in green color. Computation is done with $M=12$ orbitals. All quantities are dimensionless. }
    \label{fig1}
\end{figure}
The density $\rho(x)$ in the primary lattice $(V_2=0.0)$ for four possible Mott configurations is displayed in Fig.~\ref{fig1}. The upper panel is for unit filling, $N=9$
bosons are uniformly distributed in $S=9$ sites. The left panel is for weakly interacting Mott in deep lattice with interatomic interaction strength $\lambda=0.5$ and lattice height $V_1=20.0$. The right panel corresponds to strongly interacting Mott with $\lambda=5.0$ and lattice height $V_1 =10.0$. Both exhibit Mott localization and for an infinite system two Mott regimes are smoothly connected. However they differ in the degree of initial correlation. In the first case, Mott localization happens due to dominating effect of lattice depth; in the second case, Mott localization happens due to the dominating effect of interatomic interaction which affect the value of the Mott gap.  In both cases the density has nine well separated maxima for each boson, but due to stronger coherence in the strongly interacting Mott localization we observe different features in phase coherence in the presence of disorder. The computation is done with $M=9$ orbitals, each having $0.11 \%$ contribution for Mott configuration which results to nine-fold fragmented Mott $\vert 1,1,1,1,1,1,1,1,1 \rangle$. Repeating the simulation with $M=10$ orbitals, the last orbital exhibits insignificant contribution. 

The lower panels present density for double filling. To manage the computation and converged results, we reduce the number of lattice sites to three and increase the number of bosons to six, keeping lattice depth fixed to $V_1=10.0$. Left panel corresponds to weak interaction with $\lambda=2.0$, Mott phase is characterized by the localization of six bosons in three wells with vanishing overlap of density. We utilize $M=12$ orbitals and convergence is checked with orbital $M=13$. The many-body state is three-fold fragmented and is given by $\vert 2,2,2 \rangle$. The lowest three orbitals reach to an equal population each of $33.3\%$. The right panel     
is a fermionized Mott, when the interaction strength in the set up of Mott insulator phase is increased to $\lambda=10.0$. Onset of fermionization is visible, two strongly interacting bosons residing in the same well try to escape their spatial overlap. The characteristic dip at the center of each well mimics the onset of fermionization. The many-body state is configured as $\vert \left[1,1\right], \left[1,1\right], \left[1,1\right]\rangle$.

\section{Correlation in distorted Mott phases}
We pin each of the four Mott configurations independently by the second incommensurate lattice $(V_2)$ and try to understand the phase coherence. As the initial setups are strongly controlled by the lattice depth as well as the strength of interatomic interaction, there will be different scenario in the competition of Mott localization and localization introduced by the disorder. The double filling case is more interesting as they additionally possess initial intra-well coherence. \\

\subsection{Weakly correlated Mott in deep lattice with unit filling}

\begin{figure}
    \centering
         \includegraphics[width=0.5\textwidth]{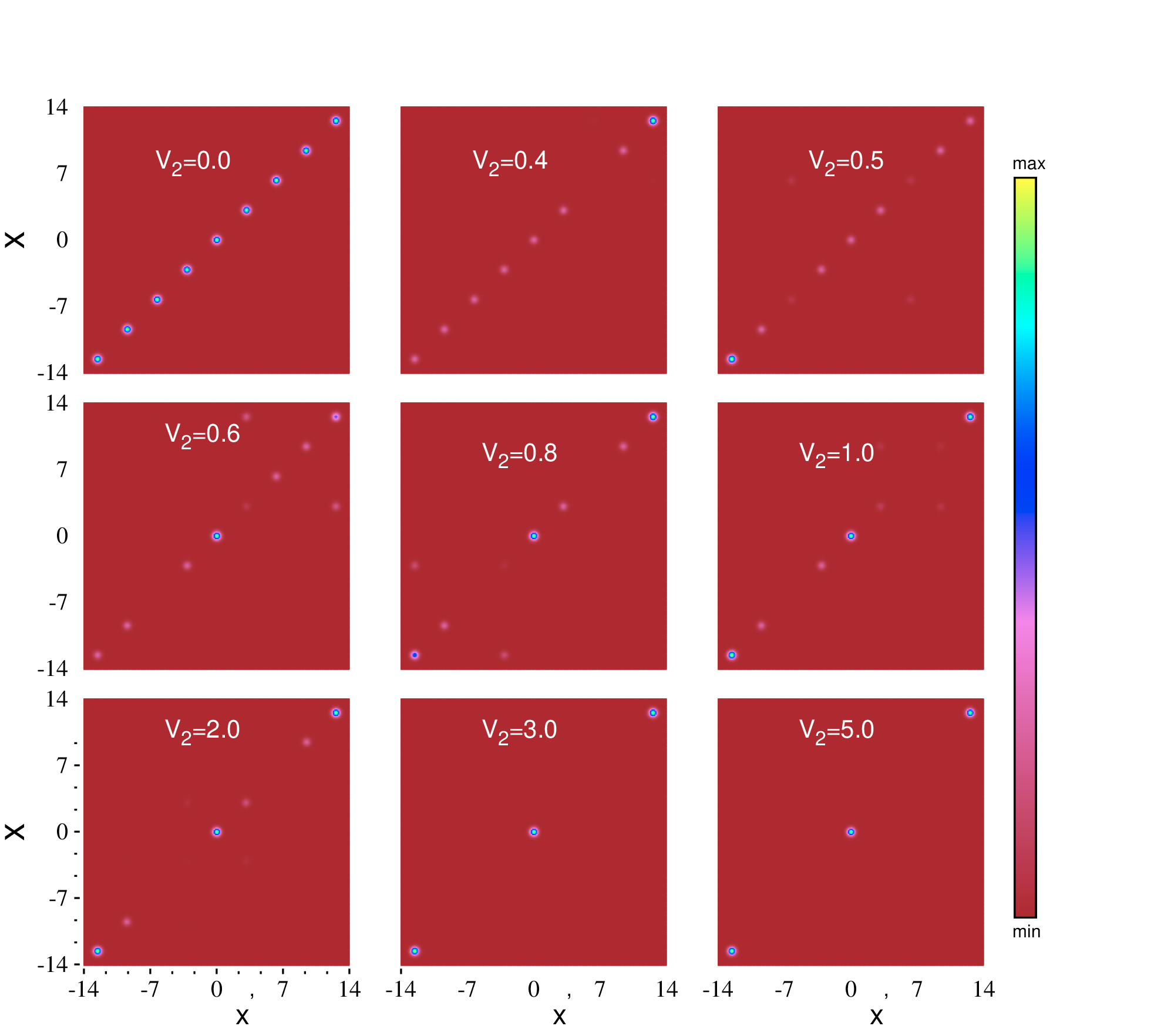}
    \caption{The reduced one-body density $\rho^{(1)}(x,x^{\prime})$ for unit filling case with $N=9$ bosons in $S=9$ lattice sites, $\lambda=0.5$, $V_1=20.0$, $M=9$ and for different disorder parameter $V_2$. All quantities are dimensionless. }
    \label{fig2}
\end{figure}

We consider the set up of top left panel in Fig.~\ref{fig1} with unit filling, $N=9$ bosons in $S=9$ lattice sites, $\lambda=0.5$, $V_1=20.0$. To explore the effect of disorder on the Mott localization we present the reduced one-body density $\rho^{(1)}(x,x^{\prime})$ for different disorder parameter in Fig.~\ref{fig2}. For $V_2=0.0$, the one-body density is perfectly diagonal, only self-correlation within each Mott peak is observed. The diagonal correlation is maintained till $V_2=0.3$. At $V_2=0.4$, the Mott localization starts to be affected due to disorder. With gradual increase in disorder, the localization due to Mott correlation and that due to disorder strongly interplay. At $V_2=2.0$, localization starts to settle in the central lattice and two outer lattice sites. At $V_2=3.0$, localization completely settles in these three sites. With further increase in disorder, localization sites do not change, only it becomes more deepened.   

It is observed in Fig.~\ref{fig2} that the competition between Mott localization and disorder leads to a distorted Mott phase. Initially, the fully coherent uniform diagonal Mott correlation in the primary lattice becomes disturbed with small disorder. Then stronger correlation is settled only in three lattice sites when the remaining six lattice sites become absolutely incoherent. We present the results for different system size: $N=5$ bosons in $S=5$ sites; $N=7$ bosons in $S=7$ sites and $N=11$ bosons in $S=11$ sites in Appendix A. We observe the same physics, both for the smaller and larger system sizes, i.e., localization due to disorder strongly competes with the Mott localization. However, for smaller system size, localization happens in the central lattice only whereas for larger system size, localization happens in the central and two corner lattice sites. 

\begin{figure}
    \centering
         \includegraphics[width=0.5\textwidth, angle=270]{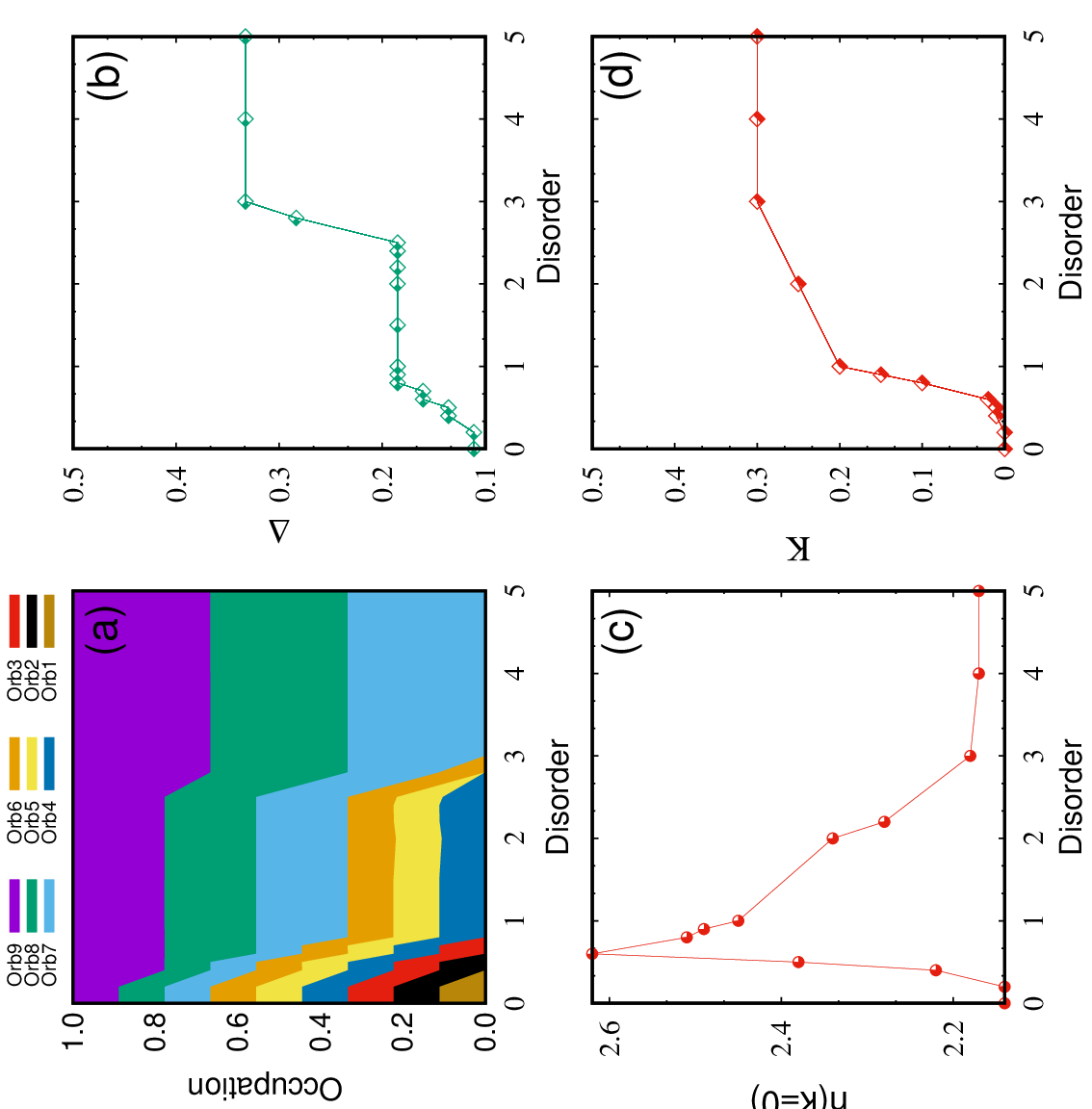}
    \caption{(a) Natural occupations(cumulative plot) as a function of disorder. (b) Dependence of the order parameter on the disorder. (c) Zero-momentum peak as a function of the disorder. (d) Dependence of the compressibility on the disorder. All quantities are dimensionless. }
    \label{fig3}
\end{figure}

However for better understanding the effect and in search whether any additional phase interrupts the phenomena, we calculate several other quantities and plot them in Fig.~\ref{fig3}. As the fragmentation is the hallmark of MCTDHB, we present natural occupations as a cumulative plot as a function of disorder strength in Fig.~\ref{fig3}(a). We find, initially the Mott state is maximally fragmented, all nine orbitals have exactly equally contributing eigenvalues. With introduction of disorder, as Mott phase looses its diagonal correlation, degree of fragmentation is gradually reduced. Number of significantly contributing orbitals is reduced and finally settled to three-fold fragmentation at $V_2=3.0$, exactly when localization happens in the central and two other outer lattice sites as seen in Fig.~\ref{fig2}. The three-fold fragmentation is maintained further with larger disorder strength. Thus the system does not enter to the superfluid phase which will be dominated by a single orbital. In Fig.~\ref{fig3}(b), we plot the order parameter as a function of disorder. Initially, for pure Mott localization, order parameter is $0.11$ which is exactly $\frac{1}{9}$. With increase in disorder, order parameter gradually increases, it indicates the deviation from Mott phase, finally at $V_2=3.0$, it settles to $0.33$. For superfluid phase it would settle to $1.0$. In Fig~\ref{fig3}(c), we plot the zero-momentum peak, $n(k=0)$ as a function of disorder. It exhibits the competition between Mott localization and localization due to disorder. The peak exhibits some irregular features, non-monotonic dependence on disorder. For smaller disorder, it first increase, which indicates that phase coherence increases due to loss of Mott correlation. With larger disorder, it decreases as disorder introduces localization.  We calculate the compressibility from fluctuation in the momentum distribution and plot it as a function of disorder in Fig.~\ref{fig3}(d). For Mott phase compressibility is zero as expected, becomes nonzero in the disordered phase which indicates the Bose glass phase. Thus all the observations conclude, the disordered Mott phase enters the Bose glass phase at a critical disorder parameter and remains as a compressible insulator.

%To further assess the correlation between atoms in the localization process we investigate the lowest orbital occupation as a function of disorder strength and plot in ~\ref{fig4} (top panel). Initially the occupation is 0.11 for the fully fragmented Mott state (without disorder). On switching on the disorder, the population in the lowest orbital slowly increases which signifies that the Mott localization is affected by disorder. When $V_2 \simeq 0.6$ lowest orbital occupation $n_1$ reaches a maximum value which signifies that Mott correlation is significantly affected by disorder. In the intermediate region, we observe a plateau which is in agreement with the observation in ~\ref{fig2}, Mott localization and localization due to disorder strongly competes. 
%

\begin{figure}
    \centering
         \includegraphics[width=0.5\textwidth, angle=270]{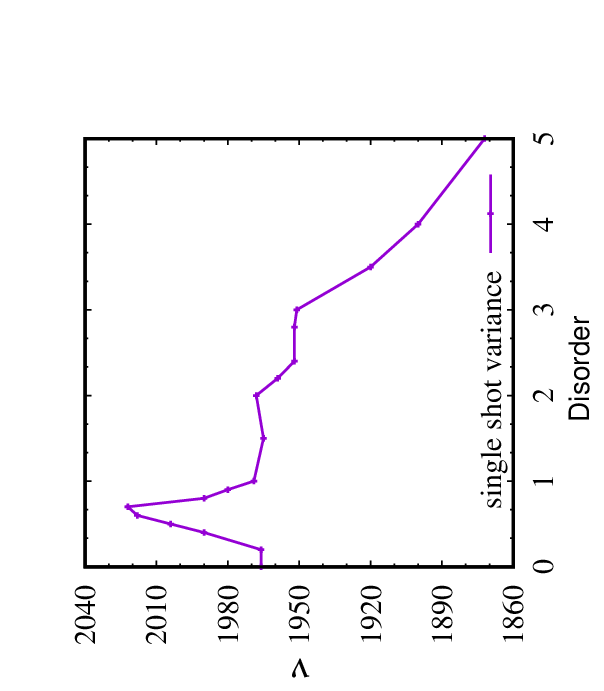}
    \caption{The variance of single shots $\nu$ in real space as a function of disorder strength. For every point, the variance of 10000 single-shot images are computed. The value is minimal for initial Mott state when Mott state localizes in nine lattice sites, it is maximum when Mott localization is strongly affected by disorder, the plateau region with few maxima and minima exhibits competition between Mott localization and localization due to disorder. It sharply decreases for larger disorder strength when localization due to disorder dominates and localization settles to only three lattice sites. All quantities are dimensionless. }
    \label{fig4}
\end{figure}

However the direct measurement of the detection of order parameter and the eigenvalues of the reduced one-body density matrix would be experimentally challenging. In the experiments, absorption image measures the positions of all particles simultaneously. Specifically the single-shot measures contains the information of inter-well and intra-well atomic correlation. We mention that fragmentation can be detected in the variance of single-shot measures both in real and momentum space ~\cite{Budhaditya,Sakmann,Bruder}. Here we use the spatial measurements because the real space single-shot distribution can be directly correlated to the localization of bosons. We adopt the procedure described in Ref.~\cite{Sakmann,Budhaditya} and calculate the spatial single-shot measure. The single-shot variance $\nu$ is plotted as a function of disorder in ~Fig.\ref{fig4}. The behaviour of single-shot variance can be understood in the following way. In the initial Mott localization (without disorder) the atoms are localized in each lattice site. As a consequence, the variance in single-shot measurement is small. We find till $V_2 = 0.3$, variance is constant when the Mott correlation is not affected by the disorder potential. Then, variance increases and around $V_2 = 0.65$, it reaches a maximum value when Mott localization is significantly affected by the disorder. When localization due to disorder starts to dominate single-shot variance starts to decrease. We also find a plateau kind region with few local maxima and minima where Mott localization and localization due to disorder competes. With further increase in disorder, for $V_2 = 3.0$, when localization due to disorder strongly dominates, the single-shot variance quickly decreases. The value is significantly smaller than that of initial Mott localization. The observation can be understood in the following way. When the initial Mott state localizes in all nine lattice sites, disorder induces localization in only three lattice sites.  Thus we find that the single-shot variance is able to detect how the Mott localization and correlation become distorted upon switching on the disorder.

\subsection{Strongly correlated Mott in intermediate lattice with unit filling}

In this section we consider the set up of strongly correlated Mott of unit filling, $N=9$ bosons distributed over $S=9$ lattice sites. The primary lattice depth is reduced to $V_1=10.0$ and interaction strength is increased to $\lambda=5.0$ to initiate a fully fragmented Mott. However as in this case the Mott localization happens due to interaction, the many-body phase is strongly correlated compared to the previous case. One can in principle make a Mott phase in a shallow lattice with larger interaction, however we face a serious problem in handling the convergence. So the present calculation deals with lattice of intermediate depth. In Fig.~1 (top right panel), the initial state is a clean Mott phase, the many-body state is maximally fragmented and is given by $\vert 1,1,1,1,1,1,1,1,1 \rangle$. We compute the density with $M=9$ orbitals, each shows occupation of $0.11$. 
\begin{figure}
    \centering
\includegraphics[width= 0.8\textwidth, angle = 270]{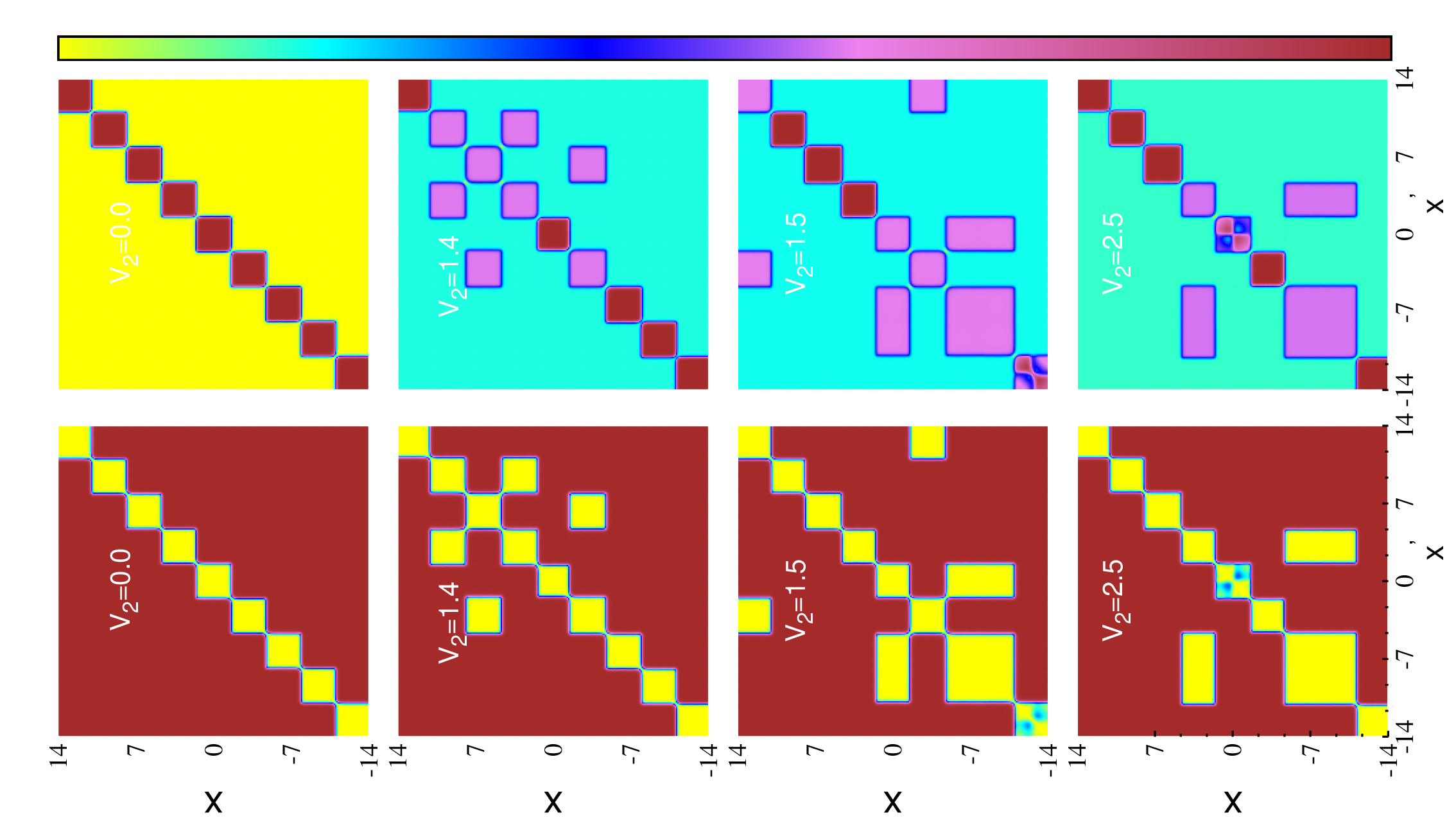}
    \caption{First and second order coherence for strongly correlated Mott with $N=9$ bosons in $S=9$ sites.
    The lattice depth of the primary lattice $V_1=10.0$, interatomic interaction $\lambda=5.0$. Parameter $V_2$, controls the disorder strength. Computation is done with $M=9$ orbitals. The left panels represent one-body and the right panels correspond to the two-body coherence. The normalized correlation function describe the three processes: leaking, melting and central localization in the lattice sites. All quantities are dimensionless. }
    \label{fig5}
\end{figure}
\begin{figure}
    \centering
    \includegraphics[width= 0.4\textwidth, angle = 270]{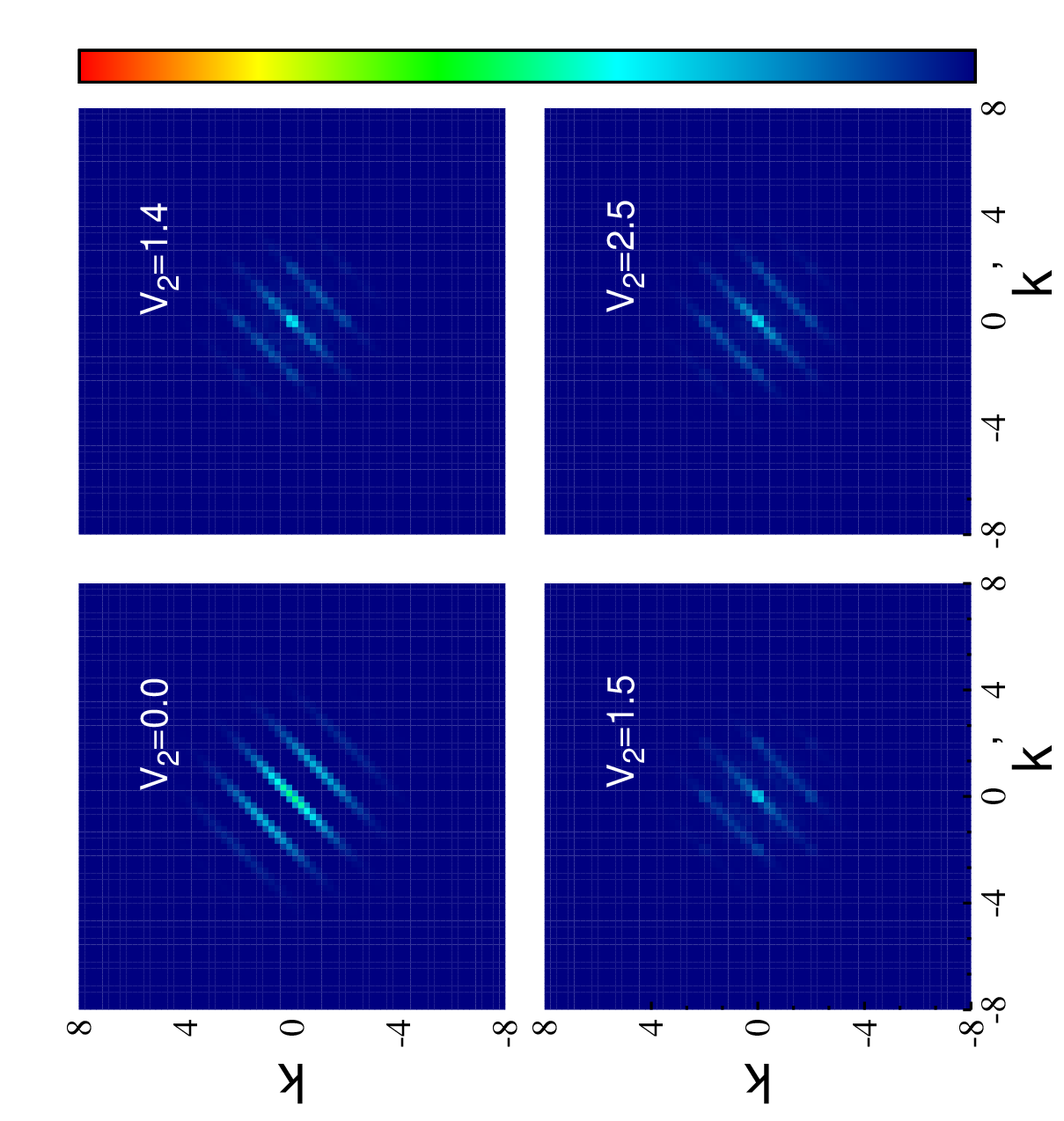}
    \caption{Effect of disorder on reduced one-body momentum distribution $\rho^{(1)}(k,k^{\prime})$ for the strongly correlated Mott with unit filling for $N=9$ bosons distributed over $S=9$ lattice sites. The primary lattice depth $V_1=10.0$, interaction strength $\lambda=5.0$ and number of orbitals $M=9$. The disorder parameter $V_2$ is tuned. The coherence zone and the central peak value characterizes how the clean Mott phase faces melting due to disorder and then the localization dominates. All quantities are dimensionless. }
    \label{fig6}
\end{figure}

Next we switch on disorder and increase its strength gradually. 
Fig.~\ref{fig5} displays the one-body and two-body correlation functions for various disorder strength $V_2$, which pins the clean Mott phase. The left panel corresponds to one-body and the right panel corresponds to two-body correlation functions. The top panel corresponds to correlation in the primary lattice $(V_2=0)$. The diagonal of the first order correlation function shows nine completely separated lobes which are highly coherent, $|g^{(1)}|^2 \simeq 1$. For maximally fragmented and fully localized Mott state off-diagonal, $(x \neq x^{\prime})$ correlation $|g^{(1)}|^2=0$. Where $g^{(2)}$ exhibits nine dark lobes along the diagonal which are completely incoherent-we refer these as {\it{correlation holes}}. The correlation holes are created due to strong {\it{antibunching effect}}. It implies that the probability of finding double occupation in a single well is practically zero. Whereas second-order coherence is completely maintained along the off diagonal.\\

We find, up to $V_2=1.3$, the system does not respond to disorder, it maintains full fragmentation, $100 \%$ one-body diagonal correlation and loss of two-body diagonal correlation.  All essential features of the clean Mott phase of the primary lattice are maintained even when the disorder parameter is significant. 
At $V_2=1.4$, we find the onset of disorder starts weakening the on-site correlation. We observe the third lattice on the right side from the central lattice exhibits some leakage in the one-body correlation. The term `leakage' means, the initially concentrated correlation in isolated Mott lobe gets a path to spread in the off-diagonal region. It signifies that disorder depletes this lattice, although we do not observe any melting in the lattice site. By the term `melting' we refer the situation when two initially separated consecutive Mott lobes become a single lobe. Natural orbital occupation shows that nine orbitals which were equally populated initially, become depleted in the presence of disorder. Five orbitals maintain initial population, each of $0.11$, i.e., five bosons occupy five orbitals with equal weight. Whereas the remaining population is redistributed unequally between the other four orbitals. This is clearly visible in the two-body correlation function, only five correlation holes survive. The remaining correlation holes are being depleted due to disorder and some minor correlation holes are additionally created around the depleted lattice site. With slight increase in disorder strength, $V_2=1.5$, we observe second and third lattice melt, one-body correlation spreads around the two lattice sites, interwell coherence is built up. Some leakage around the fourth lattice is also visible.  Out of five equally populated orbitals, two additionally become depleted.  Thus only three natural orbital maintains the population each of $0.11$. Corresponding two-body correlation function exhibits exactly three correlation holes. The correlation holes at the second and third lattice sites are joined due to disorder. We observe very strong competition between localization due to strong correlation in the lattice sites and localization due to disorder. It is clearly seen both in one-body and two-body correlation function. When disorder exhibits partial delocalization of the bosons from some selected lattice sites, some localization happens in the left most lattice site.  The corresponding fully coherent Mott lobe becomes partially incoherent and the two-body correlation exhibits appearance of partial two-body coherence in the left most lattice site. It clearly signifies that localization due to correlation and disorder compete against each other in a very complex way. We are able to present the microscopic picture of the underlying mechanism. With further increase in $V_2$, competition persists between leakage, melting and localization. Finally at $V_2=2.5$, we observe the localization due to disorder starts to dominate. One can observe the signature of localization at the central lattice in the one-body correlation. The number of natural orbitals starts to regain their population, five natural orbitals exhibit equal population of $0.11$. Number of correlation holes in the two-body correlation function increase to five, although some signature of melting persist. With further increase in disorder parameter, this competition continues; the system remains as distorted Mott phase only.\\

In the experiments, the major source of information of the disordered Mott phases can be analyzed from the momentum distribution function. In Fig.~\ref{fig6}, we plot the reduced one-body momentum distribution function $\rho^{(1)}(k,k^{\prime})$ for the different disorder strengths. In the absence of disorder, the range of coherence is wide and the momentum distribution has a weak peak at $k=0$. On switching on disorder, as the Mott regions leak and melt, phase coherence increases. The range of coherence starts to reduce, the zero momentum peak becomes stronger as seen for $V_2=1.4$ and $1.5$. For disorder $V_2=2.5$, as the localization due to disorder dominates, the coherence zone starts to spread and zero momentum peak starts to decrease. \\

In the above table (Table I), we present the results for single-shot measurements (as described before for the weak Mott states) in the real space for some selected parameter of disorder potential. It can be seen that the variance is smaller for the initial Mott localization, it increases slightly when Mott localization is distorted due to secondary lattice. With further increase in disorder strength, when localization at the central lattice happens due to dominating effect of disorder, variance sharply falls. We expect that the real-space variance of single-shot measurements will be able to distinguish the competition between Mott localization and localization due to disorder.   

We present the results for smaller system size ($N=7$ bosons in $S=7$ lattice sites) and for larger system size ($N=11$ bosons in $S=11$ lattice sites) in Appendix A. We observe the same physics as observed for $N=9$ bosons in $S=9$ lattice sites. However, we note some finite size effect, larger system claims larger disorder strength for central localization, although the basic observations of melting of Mott lobes and the leakage of diagonal coherence in the off-diagonal region remain same. We also mention the comparison of our findings with the recent work on disorder induced delocalization for Lieb-Liniger gas in shallow quasi-periodic potential~\cite{yao2024motttransitionlieblinigergas}. Considering the phase diagram,  Fig. 1(b) of Ref.~\cite{yao2024motttransitionlieblinigergas}, the strongly interacting Mott in the periodic lattice remains as Mott in the quasi-periodic lattice, which apparently contradicts our observations. We like to point out that the observation of distortion in Mott correlation in our present work may be the typical effect of finite systems which may not be observed for truly large system size. Or, the usual fingerprints like superfluid order parameter and the compressibility to distinguish the BG phase from the Mott phase are inadequate to detect the distorted Mott localization. We stress the in situ correlation measures~\cite{Schweigler} can distinguish the distorted Mott from the true Mott.

\begin{table}
\centering
\begin{tabular}{ || c | c || }
\hline \hline
disorder ($V_2$) & shingle-shot variance ($\nu$)  \\
\hline \hline
0.0 & 2283 \\
\hline
1.4 & 2293 \\
\hline
1.5 & 2089 \\
\hline
2.5  & 2081 \\
\hline \hline
\end{tabular}
\caption{Single-shot variance for different disorder parameter (10000 samples per data point). The system is $N=9$ bosons in $S=9$ lattice sites. The lattice depth of the primary lattice $V_1$ = 10.0, interatomic interaction $\lambda=5.0$. Computation is done with $M=9$ orbitals. All quantities are dimensionless.} 
\end{table}

\subsection{Weakly interacting and fermionized Mott with double filling}

\begin{figure}[ht!]
    \centering
    \includegraphics[width= 0.8\textwidth, angle = 270]{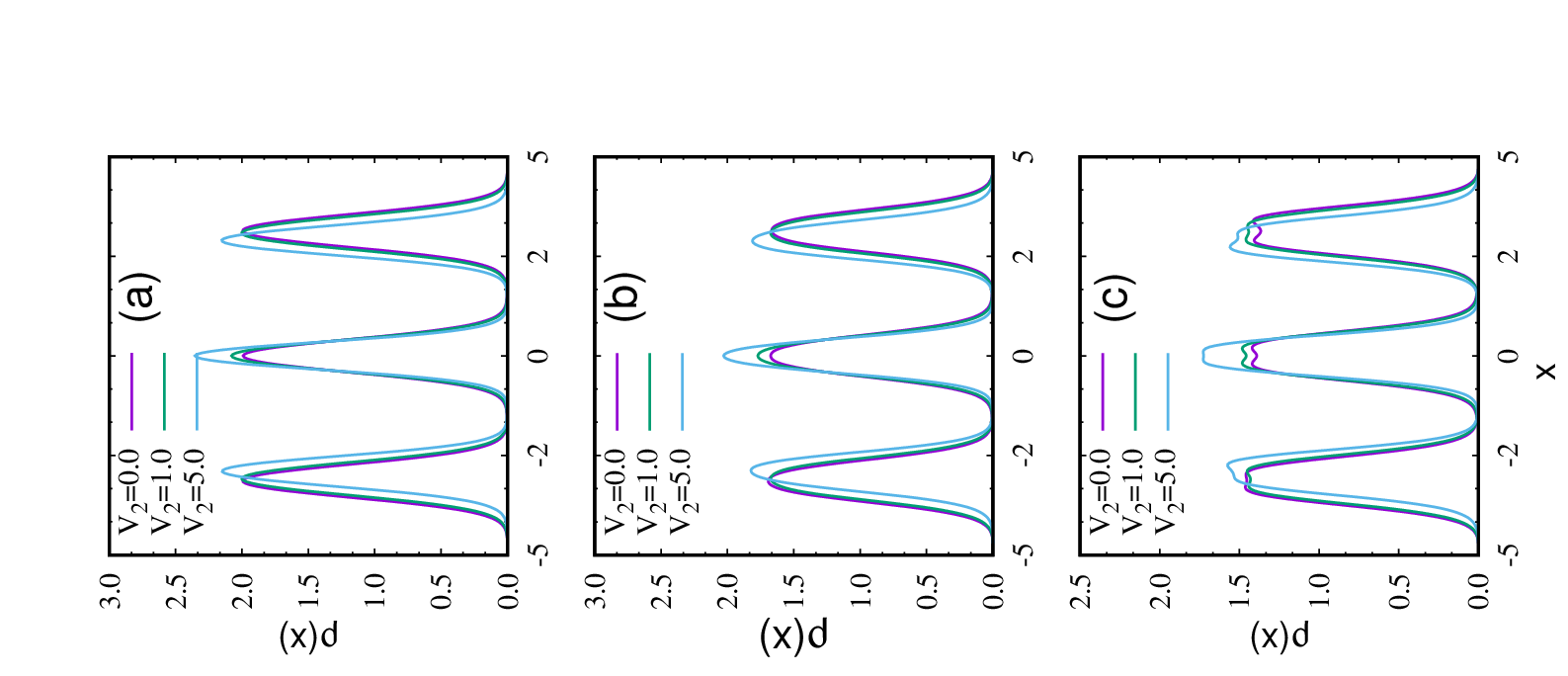}
    \caption{The one-body density $\rho(x)$ for the Mott insulator and fermionized Mott phase with double filling in disordered lattice. $N=6$ bosons are in $S=3$ sites of primary lattice of depth $V_1=10.0$. The top panel (a), corresponds to Mott phase with interatomic interaction $\lambda=2.0$, The disorder potential introduces stronger central localization only. The middle panel (b), is also a Mott phase with higher interaction potential $\lambda=5.0$. The disorder interferes the Mott localization in each lattice. The bottom panel (c), corresponds to $\lambda=10.0$ which represents a fermionized Mott phase with clear signature of on-site repulsion between two strongly interacting bosons. The disorder potential now competes with the on-site interaction and the 'characteristic dip' in each lattice gradually disappears. The calculation is done with $M=12$ orbitals. All quantities are dimensionless. }
    \label{fig7}
\end{figure}

 For filling factor more than one, the localization process and the effect of disorder is more intriguing and less understood. The on-site repulsion would cause the fragmentation of particles which is beyond the validity of Bose-Hubbard model.
Fig.~\ref{fig7} demonstrates the case of filling factor two, six particles in three wells, the on-site interaction effects become prominent due to higher density of particles. We plot the one-body density for different choices of two-body interaction strength. We aim to observe the response of the system when the secondary lattice is introduced. For the choice of on-site interaction $\lambda=2.0$ (top panel of Fig.~\ref{fig7}), the one-body density exhibits three peaks in three wells of the primary lattice. In terms of Bose Hubbard model, we have a formation of a "Mott state" of two particles per site. The existence of on-site repulsion is not visible for this choice of $\lambda$. Upon introducing the secondary lattice, the peak in the central lattice becomes more prominent, but its position remains unaltered. Whereas the two peaks in the left and right wells become stronger as well as slightly pushed towards the center. We observe the same physics even when $\lambda=5.0$ (middle panel of Fig.~\ref{fig7}). However for $\lambda=10.0$ (bottom panel of Fig.~\ref{fig7}), the on-site interaction is clearly visible, fermionization happens and the observed effect goes beyond the Bose-Hubbard model. On introducing the second lattice, the "characteristic dip" in the central well starts to disappear, which signifies that disorder now competes to the on-site repulsion. The effect in the outer lattice is less although the characteristic dips start to be distorted gradually.

\begin{figure}
    \centering
    \includegraphics[width= 0.4\textwidth, angle = 270]{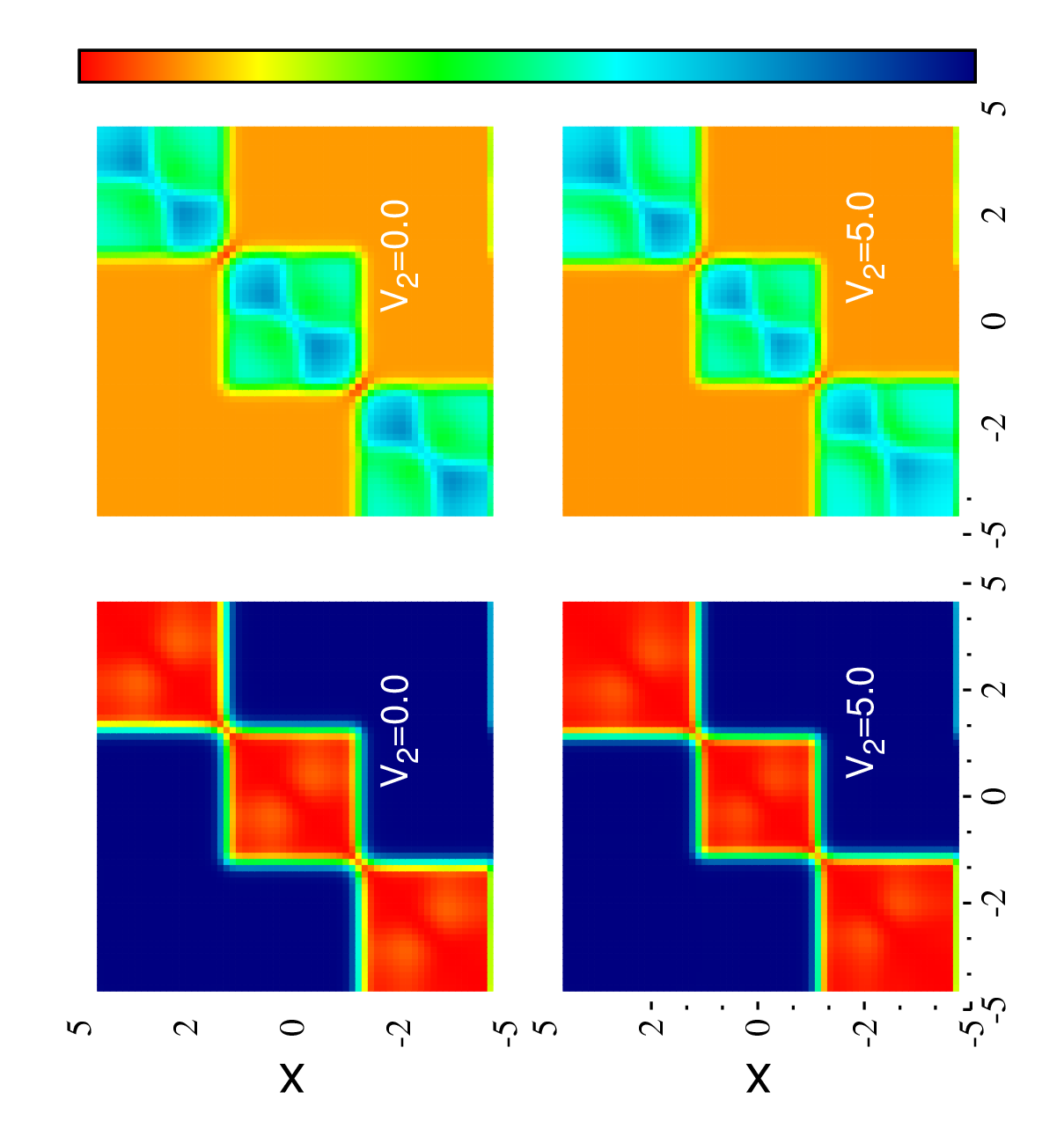}
    \caption{First and second order coherence for weakly correlated Mott for dual filling, with $N=6$ bosons in $S=3$ sites. The lattice depth of the primary lattice $V_1=10.0$, interatomic interaction $\lambda=2.0$, disorder parameter $V_2$ is increased gradually. Computation is done with $M=12$ orbitals. The left panel represents one-body and the right panel corresponds to the two-body coherence. The internal structure in both one- and two-body correlation function describes two interacting bosons in each lattice sites.  Due to additional intra-well correlation, it is very hard to pin the lattice sites with a secondary lattice. For very strong disorder $V_2=5.0$, we observe the central localization becomes stronger.  All quantities are dimensionless. }
    \label{fig8}
\end{figure}

\begin{figure}
    \centering
    \includegraphics[width= 0.4\textwidth, angle = 270]{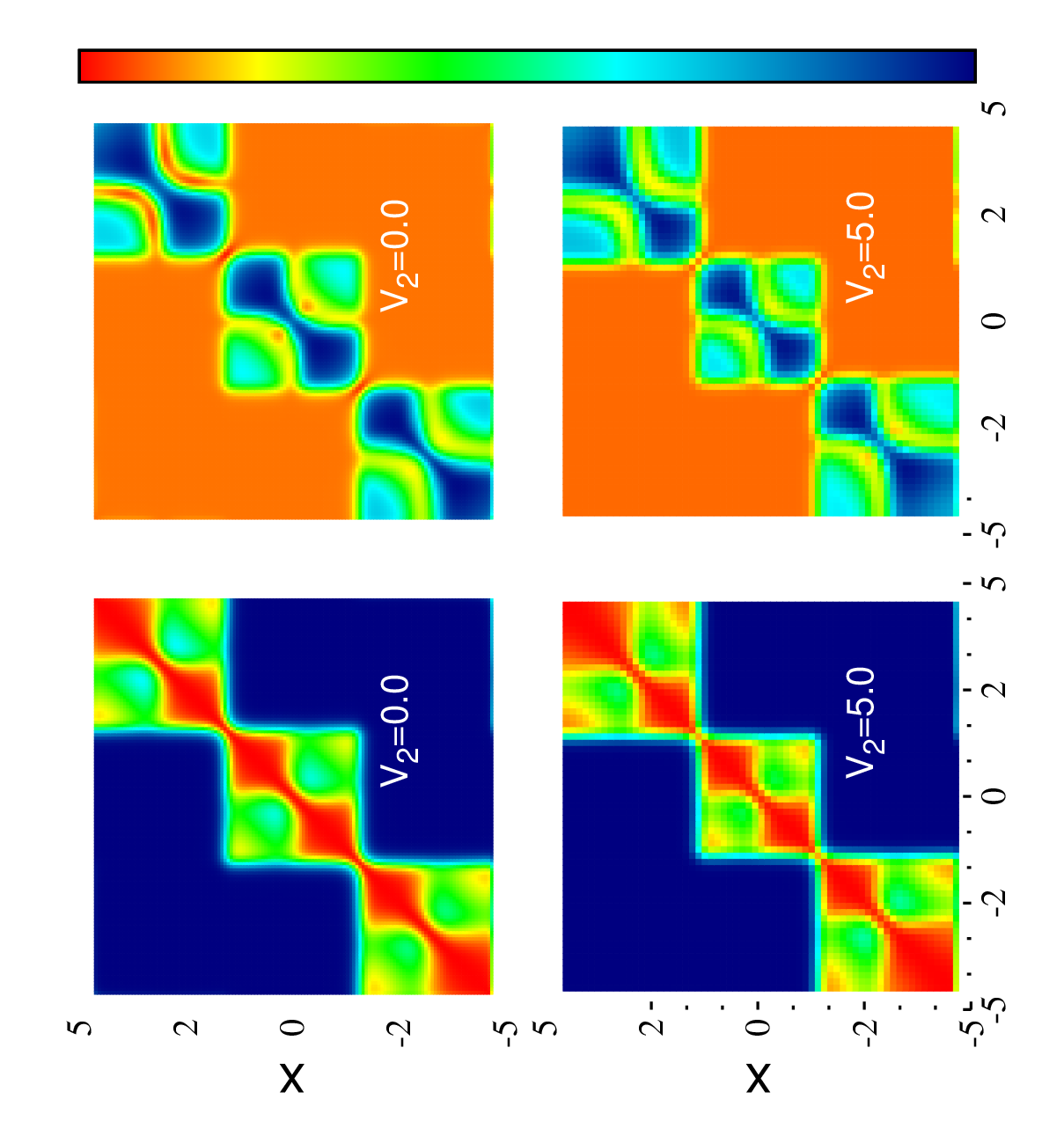}
    \caption{ First and second order coherence for the set up of fermionized Mott in the primary lattice. $N=6$ very strongly interacting bosons with $\lambda=10.0$ are localized in $S=3$ lattice sites of lattice depth $V_1=10.0$. Computation is done with $M=12$ orbitals. The internal structure becomes prominent, each well exhibit two separated lobes signifying that two strongly interacting bosons avoid their spatial overlap. The two-body correlation also shows a complete isolation of two particles in each well. The secondary lattice distorts the on-site correlation. All quantities are dimensionless.} 
    \label{fig9}
\end{figure}

%\begin{figure}
%    \centering
%    \includegraphics[width= 0.4\textwidth, angle = 270]{Fig8.png}
%    \caption{Effect of disorder on reduced one-body momentum distribution $\rho^{(1)}(k,k^{\prime})$ for the strongly correlated Mott with double filling for %$N=6$ bosons distributed over $S=3$ lattice sites. The primary lattice depth is $V_1=10.0$ and the number of orbitals is $M=12$. The disorder parameter $V_2$ is tuned. Top panel ( a and b) corresponds to weak interaction, $\lambda=2.0$, bottom panel ( c and d) corresponds to strong interaction, $\lambda=20$. The coherence zone and the central peak characterizes the effect of disorder for weak Mott. The momentum correlation remains intact for fermionized Mott even when the disorder is very strong.  All quantities are dimensionless. }
%    \label{fig9}
%\end{figure}

Fig~\ref{fig8} presents the one- and two-body correlation functions for the weakly interacting Mott with double filling. The primary lattice height is kept $V_1=10.0$ and interaction strength is chosen $\lambda=2.0$. Due to filling factor two, the intra-well structure is developed both in one- and two-body correlation function. The presence of two interacting bosons in each lattice is visible, however as the interaction is not too strong, the two bosons can not completely escape their spatial overlap. Our further intention is to examine whether the additional disorder can melt the lattice or not. We find that the strongly correlated Mott remains silent even for very high disorder strength. When the disorder is increased to $V_2=5.0$, central localization becomes stronger only. For much higher disorder strength (not shown here), inter-well coherence between populated neighbouring wells is gradually developed. We specifically observe the restoration of coherence between the central and the left well. The internal structure in the left well also becomes distorted due to disorder. We  observe some change in two-body coherence in the central lattice whereas the coherence in the right most lattice remains as before. \\  

In Fig.~\ref{fig9}, we plot the correlation function for very strong interaction in the fermionized limit, when interaction strength is increased to $\lambda=10.0$. We consider the same set up of dual filling otherwise. Comparing Fig~\ref{fig8}, we find distinct intra-well structure. Two distinct bright lobes signify the presence of two fermionized bosons. The bosons in each site are now spatially separated dimer. The coherence between wells is already lost as $|g^{(1)}|^2$ $\simeq 0$ for all off diagonal points. Complete spatial isolation in each lattice mimics the onset of fermionization. We find that the secondary lattice distorts the on-site correlation only.

\section{Conclusion}
Let us briefly summarize the main results of the paper. We have computed the consequences of disorder on the Mott insulating phases in 1D optical lattice. Although it is an established fact that Mott to superfluid transition in disordered lattice is intercepted by intermediate Bose glass phase, the corresponding study of phase coherence is highly limited~\cite{Pinaki:NJP}. In the present work, we employ multiconfigurational time dependent Hartree for bosons, which extends the physics beyond the Bose-Hubbard model. Controlling the lattice depth, inter atomic interaction and filling factor, we are able to construct four emergent Mott phases in the primary lattice. All the four phases are distinctly different by their correlation in the primary lattice.  
%We try to understand the %microscopic picture how %the disorder pin the %individual lattice sites %for these correlated %many-body phases. 
Mott phase in deep lattice with unit filling factor is less correlated and emerges as a Bose glass phase in the presence of disorder. However strongly correlated Mott achieved for strongly interacting bosons localized in an intermediate lattice exhibits fascinating features.  We find very complex competition between Mott correlation and localization due to disorder which is manifested as a signature of Mott lobe's opening, melting of Mott lobes and central localization. 

Whereas the Mott insulator phase with filling factor two possess intra-well coherence and exhibits more intriguing physics. The pair of bosons in each lattice can be made weakly interacting as well as strongly interacting. The Mott phase in the first case is flexible, both inter-well and intra-well coherence are affected by the secondary lattice. Whereas in the fermionized limit, Mott exhibits fragmentation in each lattice. With strong disorder, the onset of fermionization is gradually lost. Our in-depth analysis exhibits how the tailored disorder can be utilized to control the intra- and inter-well correlation. 

The finite-size ensemble considered in the present work does not exhibit the true macroscopic systems in strict sense. However, the ground state properties observed for small ensemble are analogous to those for macroscopic systems. Thus in our present work finite-size effect prevails but that does not dictate the observed physics. The different Mott phases considered in this article can be obtained by appropriate tuning of the lattice depth, interaction strength and filling factor in the set up of 1D optical lattice in typical cold-atom experiments. Our extensive study of coherence in real space is supplemented by the one-body momentum distribution and the variance of spatial single-shot variance, which can be observed in the matter wave interference. Experimental absorption images in the momentum space may also be expected to contain and infer the information about the correlation of atoms. Investigation into incommensurate filling and the phase correlation of strongly interacting dipolar bosons are possible extensions of this work. 

\section*{Acknowledgments} 
BC acknowledges significant discussion of results with P. Molignini,  Sk. Noor Nabi, S. Basu and P. Sengupta.
LS acknowledges the BIRD Project “Ultracold atoms in curved geometries” of the University
of Padova. LS is  partially supported by the European Union-NextGenerationEU within 
the National Center for HPC, Big Data and Quantum Computing 
[Project No. CN00000013, CN1 Spoke 10: Quantum Computing] and by the European Quantum Flagship Project PASQuanS 2. L.S. 
acknowledges Iniziativa Specifica Quantum of Istituto Nazionale di Fisica Nucleare, the Project ``Frontiere Quantistiche" within the 2023 funding programme  `Dipartimenti di Eccellenza' of the Italian Ministry for Universities and Research, and the PRIN 2022 Project ``Quantum Atomic Mixtures: Droplets, Topological Structures, and Vortices".
BC and AG thank Fundação de Amparo à Pesquisa do Estado de São Paulo (FAPESP), grant nr.~2023/06550-4. AG also thanks the funding from Conselho Nacional de Desenvolvimento Científico e Tecnológico (CNPq), grant nr.~306219/2022-0. 

%%%%%%%%%%%%%%%%%%%%%%%%%%%%%%%%%%%%%%%%%%%%%%%5

%\usepackage{appendix}
\appendix
\section{Results for different system size}
\begin{figure}
    \centering
         \includegraphics[width=0.4\textwidth, angle=270]{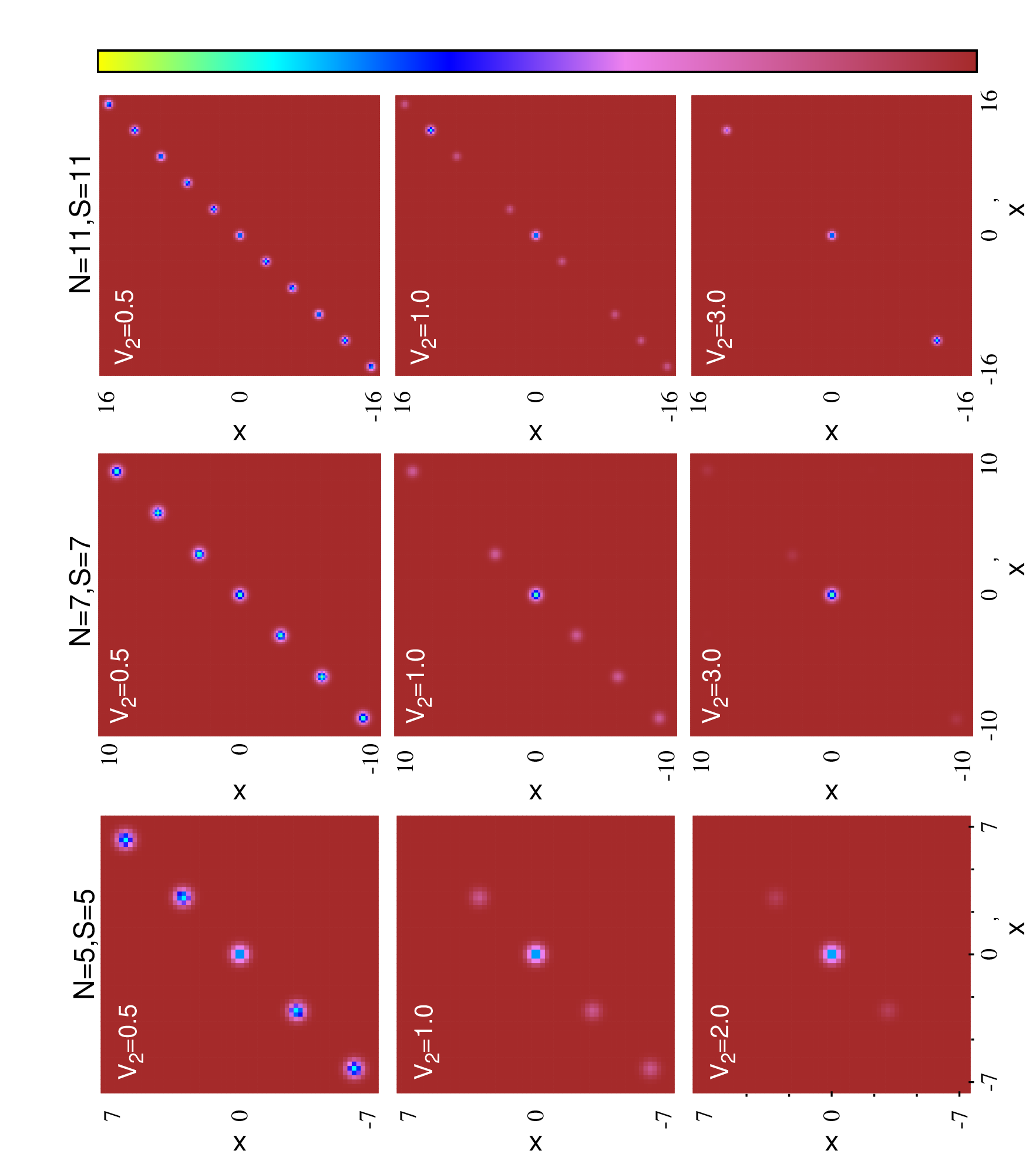}
    \caption{The reduced one-body density $\rho^{(1)}(x,x^{\prime})$ for weakly correlated Mott unit filling case with various system sizes. The left column is for $N=5$ bosons in $S=5$ lattice sites, computation is done with with $M=5$ orbitals. The middle column corresponds to $N=7$ bosons in $S=7$ lattice sites and $M=7$. The right column corresponds to $N=11$ bosons in $S=11$ lattice sites and $M=11$. For each case, convergence is checked by repeating the computation with one extra orbital. For all cases the primary lattice depth is fixed to $V_1=20.0$ and the interaction strength is fixed to $\lambda=0.5$, disorder parameter $V_2$ is varied. All quantities are dimensionless. }
   \label{fig10}
\end{figure}

In this Appendix, we show the Fig.~\ref{fig2} and Fig.~\ref{fig4} results of the main text for smaller as well as larger lattice size with unit filling factor to demonstrate that the results presented in the main text can be generalized. Similar to the main text results for $N=9$ bosons in $S=9$ lattice sites, we present the reduced one-body density $\rho^{(1)}(x,x^{\prime})$ in Fig.~\ref{fig10}  with $N=5$ bosons in $S=5$ sites; $N=7$ bosons in $S=7$ sites and $N=11$ bosons in $S=11$ sites. Following the main text, the primary lattice depth parameter is kept at $V_1=20.0$, two-body interaction strength is fixed to $\lambda=0.5$ and the secondary lattice depth parameter $V_2$ is varied. Thus Fig.~\ref{fig10} represents the case of weak Mott in deep lattice for various finite bosonic ensemble. Our general observation, the initial Mott localization is distorted due to additional secondary lattice and localization in the central and corner lattices again set up when disorder strength is higher. As expected, finite size effect is there but that did not dictate the physics. With same disorder strength, for smaller ensemble ($N=5$ bosons in $S=5$ lattice sites and $N=7$ bosons in $S=7$ lattice sites) we observe central localization happens, whereas for larger ensemble ($N=9$ bosons in $S=9$ lattice and $N=11$ bosons in $S=11$ lattice sites) localization happens in the central and two corner lattice sites. We also compute the order parameter $\bigtriangleup$ for different ensemble size and same disorder strength parameter. We find that for smaller lattice size, when quick localization in central lattice happens, the system makes transition close to superfluid, one natural orbital becomes significantly occupied. However for larger lattice, as localization happens to central as well as at the corner lattice sites, the system reaches some intermediate phase when three natural orbitals exhibit equal populations. We conclude that the observed competition between the Mott localization and localization due to disorder can be generalized for larger lattice size.  

\begin{figure}
    \centering
\includegraphics[width= 0.6\textwidth, angle = 270]{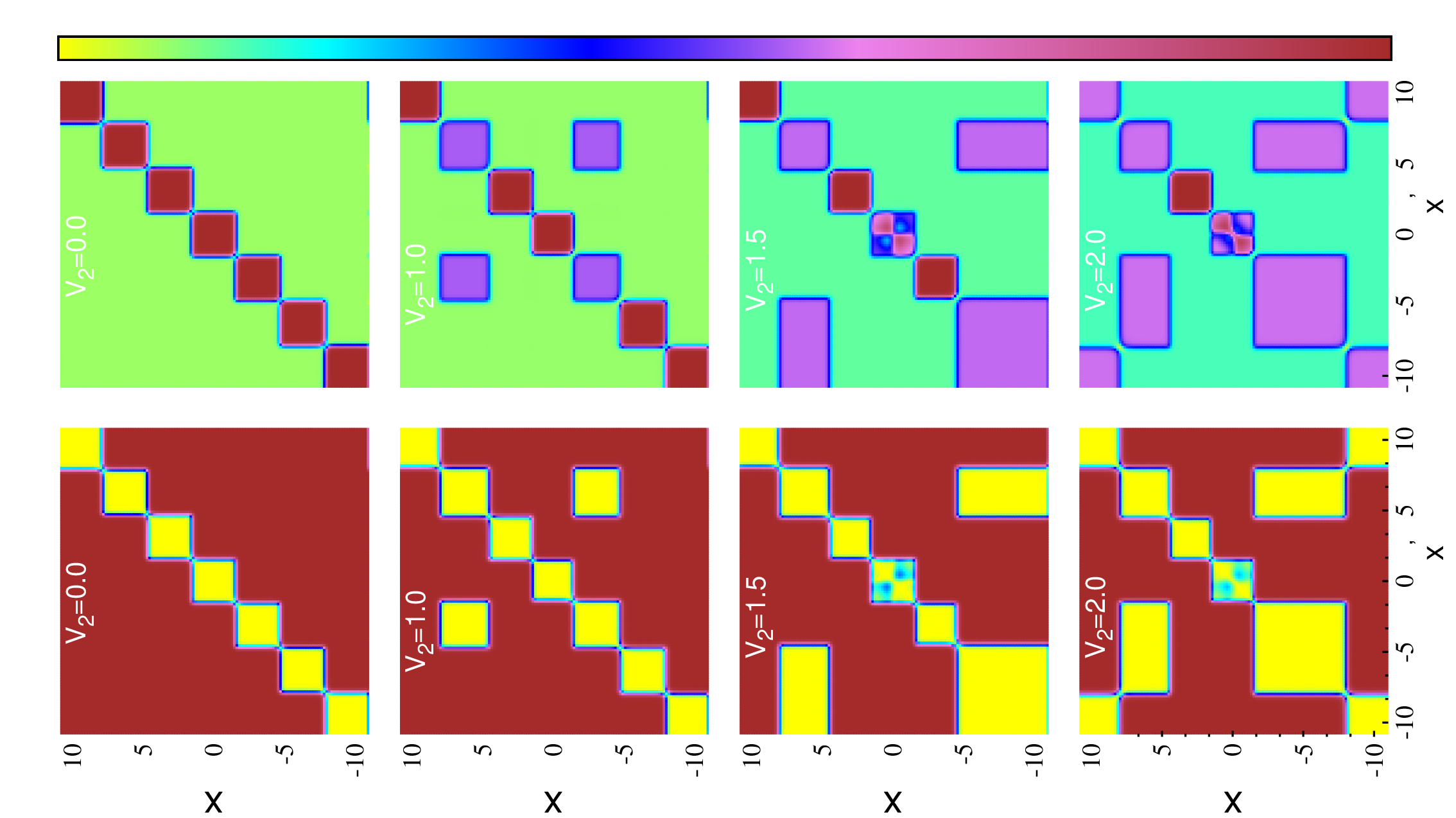}
    \caption{First and second order coherence for strongly correlated Mott with $N=7$ bosons in $S=7$ sites.
    The lattice depth of the primary lattice $V_1=10.0$, interatomic interaction $\lambda=5.0$. Parameter $V_2$, controls the disorder strength. Computation is done with $M=7$ orbitals. The left panel represents one-body and the right panel corresponds to the two-body coherence. The normalized correlation function describe the three processes: leaking, melting and central localization in the lattice sites. All quantities are dimensionless. }
    \label{fig11}
\end{figure}

\begin{figure}
    \centering
\includegraphics[width= 0.6\textwidth, angle = 270]{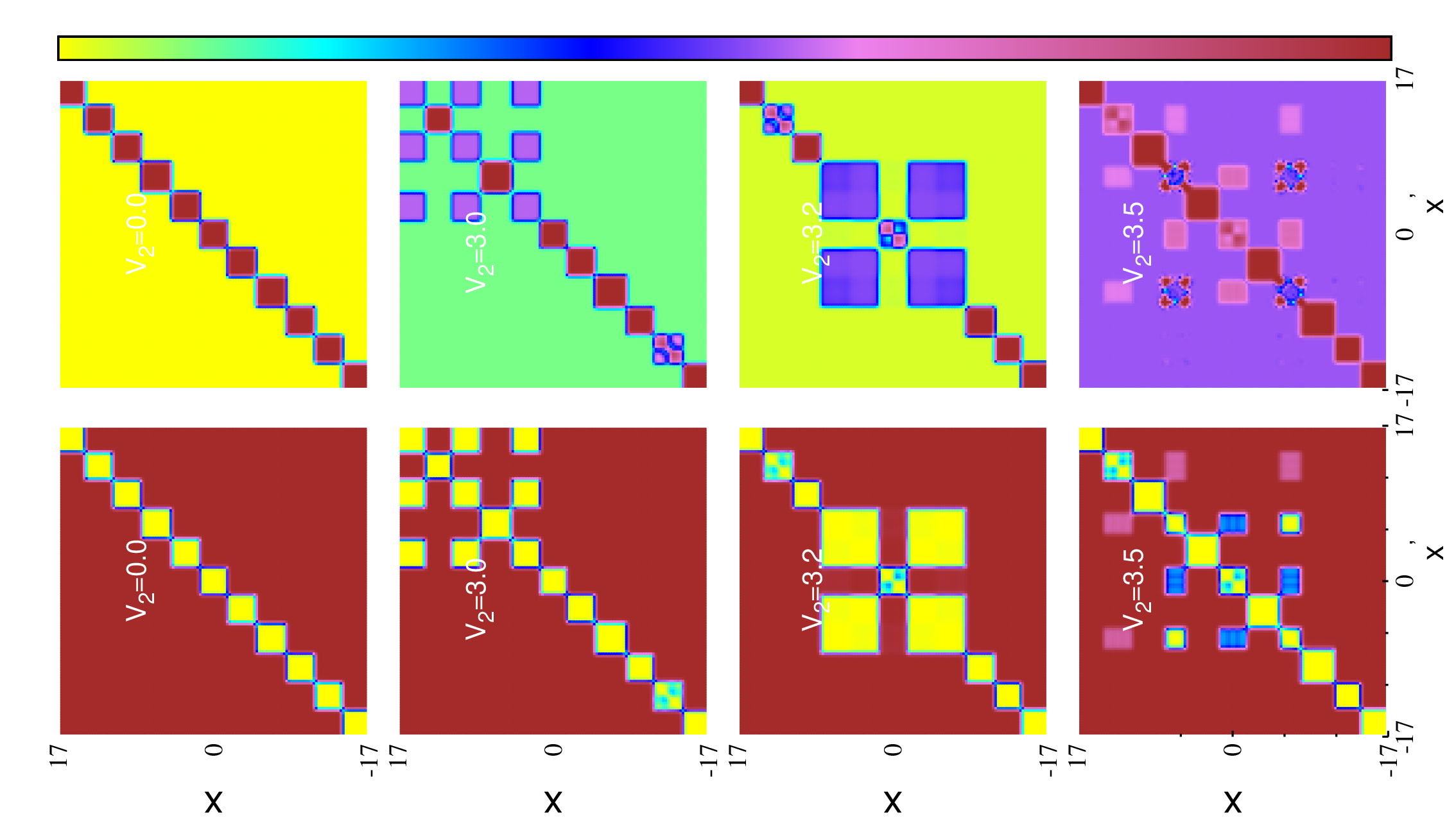}
    \caption{First and second order coherence for strongly correlated Mott with $N=11$ bosons in $S=11$ sites.
    The lattice depth of the primary lattice $V_1=10.0$, interatomic interaction $\lambda=5.0$. Parameter $V_2$, controls the disorder strength. Computation is done with $M=11$ orbitals. The left panel represents one-body and the right panel corresponds to the two-body coherence. The normalized correlation function describe the three processes: leaking, melting and central localization in the lattice sites. All quantities are dimensionless. }
    \label{fig12}
\end{figure}

In Fig.~\ref{fig11} and Fig.~\ref{fig12}, we plot the phase correlation of strongly interacting Mott in intermediate lattice depth for $N=7$ bosons in $S=7$ lattice sites and $N=11$ bosons in $S=11$ lattice sites. Following Fig.~\ref{fig4}, we keep the two-body interaction strength $\lambda=5.0$ and primary lattice depth $V_1=10.0$. Fig.~\ref{fig11} and Fig.\ref{fig12} generalize the observations made in Fig.~\ref{fig4} for $N=9$ bosons and $S=9$ lattice sites. The left column corresponds to one-body correlation function and the right column corresponds to two-body correlation function. The observed physics with $N=9$ bosons in $S=9$ lattice sites remain valid both for smaller lattice system as well as larger lattice system. For $N=7$ bosons in $S=7$ lattice sites, the seven bright lobes along the diagonal in the one-body correlation and seven correlation holes along the diagonal of two-body correlation clearly demonstrate fully fragmented many-body phase with the number configuration $\vert 1,1,1,1,1,1,1 \rangle$ in the primary lattice. On switching on the disorder, we observe strong competition between localization due to strongly correlated Mott and localization introduced by disorder. It results to melting of Mott lobes and then central localization. We observe the same physics for $N=11$ bosons in $S=11$ lattice sites as observed for $N=9$ bosons in $S=9$ lattices, but with bit higher disorder strength. Thus the observations made in the main text that the response of Mott localization due to stronger interaction is different from Mott localization due to deep lattice can also be generalized for larger lattice systems.

\section{System parameters}
In this appendix we discuss the parameters for the numerical simulations in the main text. We perform the simulations of $N=9$ bosons with $M=9$ orbitals and $N=6$ bosons with $M=12$ orbitals. The quasi periodic lattice is the superposition of two optical lattices; a primary lattice of depth $V_1$ and wavelength $\lambda_1$ and a secondary lattice of depth $V_2$ and wavelength $\lambda_2$ parameterized as 
\begin{equation}
V(x) = V_1\sin^2(k_1x) + V_2 \sin^2(k_2x)
\end{equation}
where $k_i$ is the wave vector. We choose $\lambda_1 \simeq 1032$~nm and $\lambda_2 \simeq 862$~nm, which are compatible with real experimental realizations in ultracold atomic gases. These give vectors $k_1 \simeq 6.088 \times 10^{6}$ m$^{-1}$ and $k_2 \simeq 7.289 \times 10^6$ m$^{-1}$. We set hard-wall boundaries to restrict the optical lattice to the center-most nine minima for unit filling case and three minima for double filing case. 

%%%%%%%%%%%%
\begin{table}[ht!]
\centering
\begin{tabular}{ || c | c || }
\hline \hline
Quantity & MCTDH-X units  \\
\hline \hline
unit of length &  $\bar{L} = \lambda_1/3 = 344$ nm \\
\hline
unit of energy & $\bar{E} = \frac{\hbar^2}{2 m \bar{L}^2} =E_r (\frac{3}{\pi})^2$\\
\hline
potential depth & $V=10.0 \bar{E} \approx \: 9.128 E_r$ \\
\hline
on-site repulsion & $\lambda = 0.5 \bar{E} \approx \: 0.456 E_r$ \\
\hline \hline
\end{tabular}
\caption{Units used in MCTDH-X simulations. $E_r=\frac{\hbar^2 k_1^2}{2m}$ is the recoil energy.}
\end{table}

\subsection{Lengths}
In MCTDH-X simulations, we choose to set the unit of length $\bar{L} \equiv \frac{\lambda_1}{3}= 344$ nm, which makes the minima of the primary lattice appear at integer values in dimensionless units, while the maxima are located at half integer values. $x=0$ is the center of the lattice which  can host an odd number of lattice sites $S$. In our numerical simulation, we consider an integer filling of $N=9$ bosons in $S=9$ sites and a double filling with $N=6$ bosons in $S=3$ sites. For the first case, we run simulations with 1024 grid points and 512 grid points for the second case. \\

\subsection{Energies}

The unit of energy $\bar{E}$ is defined in terms of the recoil energy of the primary lattice, i.e. $E_r \equiv \frac{\hbar^2 k_1^2}{2m} \simeq  3.182 \times 10^{-26}$ J  with $m$ $\simeq$ 38.963 Da, the mass of $^{39}$K atoms.  Thus we define the unit of energy as $\bar{E} \equiv \frac{ \hbar^2} {2m L^2}$ = $ E_r (\frac{3}{\pi})^2$ = $ 2.904 \times 10^{-26}$ J. In typical experiments with quasi periodic optical lattice the depth of the primary lattice is varied in the around few tens of  recoil energies and the depth of the secondary lattice is varied around few recoil energy. In our simulations, we probe similar regimes: $V_1$ $\in$ $\left[ 10 E_r, 20 E_r\right]$ and $V_2$ $\in$ $\left[ 0 E_r, 3 E_r \right]$. The on-site interactions are kept fixed $\lambda=0.5 E_r$ for weakly interacting Mott and $\lambda=5.0 E_r$ for strongly interacting Mott, these values can be achieved in the ultracold quantum simulators.

\subsection{Time}
The unit of time is defined from the unit of length as $\bar{t}$  $\equiv \frac{2 m \bar{L}^2}{\hbar}$ = $\frac{2 m {\lambda_1}^2}{\hbar}$= $0.1307$ $10^{-4}$ s. = $ 13.07 \mu s$.

\bibliography{ref}

\end{document}